\documentclass[]{spie}  
 
\usepackage{amsmath,amsfonts,amssymb}
\usepackage{graphicx}
\usepackage[colorlinks=true, allcolors=blue]{hyperref}
\usepackage{amsfonts}
\usepackage{physics}
\usepackage{url}
\usepackage{physics}
\usepackage{cleveref}
\usepackage{caption}
\usepackage{subcaption}
\usepackage{multirow}
\usepackage{acronym}
\usepackage{blindtext}

\title{Segment-level thermal sensitivity analysis for exo-Earth imaging}

\author[a]{Ananya Sahoo}
\author[b]{Iva Laginja}
\author[a]{Laurent Pueyo}
\author[a]{R\'emi Soummer}
\author[c]{Laura E. Coyle}
\author[c]{J. Scott Knight}
\author[d]{Matthew East}
\affil[a]{Space Telescope Science Institute, 3700 San Martin Drive, Baltimore, USA}
\affil[b]{LESIA, Observatoire de Paris, Universit\'{e} PSL, Sorbonne Universit\'{e}, Universit\'{e} Paris Cit\'{e}, CNRS, 92195 Meudon, France}
\affil[c]{Ball Aerospace \& Technologies Corp, 1600 Commerce Street, Boulder, USA}
\affil[d]{L3Harris Technologies, Inc, 332 Initiative Dr, Rochester, USA}

\begin{document} 
\maketitle

\begin{abstract}
We present a segment-level wavefront stability error budget for space telescopes essential for exoplanet detection. We use a detailed finite element model to relate the temperature gradient at the location of the primary mirror to wavefront variations on each of the segment. We apply the PASTIS sensitivity model forward approach to allocate static tolerances in physical units for each segment, and transfer these tolerances to the temporal domain via a model of the WFS\&C architecture in combination with a Zernike phase sensor and science camera. We finally estimate the close-loop variance and limiting contrast for the segments’ thermo-mechanical modes.
\end{abstract}

\keywords{Segmented telescope, thermal tolerancing, wavefront control and sensing}

\section{INTRODUCTION}
\label{sec:intro}  
Imaging and characterizing Earth-like planets is one of the key science goals identified by the NASA 2020 Astrophysics Decadal Survey\cite{2020decadalsurvey}. These habitable worlds are typically located at a small angular separation ($\sim0.1$ arcsecond) from their host star and to resolve them we need telescopes with a large primary diameter \cite{traub2010direct}. In addition to the close angular separation, these are extremely faint bodies (i.e., $\sim10^{10}$ times fainter than their host star) and high-contrast imaging (HCI) techniques prove to be useful in imaging these planets overshadowed by the bright glare of their host star. Previous studies quantify a dark hole (DH) contrast of at least $10^{-11}$ needs to be achieved at a small inner working angle of the coronagraphic point-spread function (PSF) to maximize the yield of exo-earths. For this, wavefront phase aberrations of the order $10^{-12}$ m need to be controlled \cite{lyon2012space,coyle2019large, nemati2017effects}. Future large space telescopes such as the Large UV Optical Infrared Surveyor (LUVOIR)\cite{luvoir2019luvoir,bolcar2017large} will include a segmented primary mirror to optimize their mission cost, scale, mass and possible launch vehicles. The presence of segment gaps, phasing errors between segments introduce complex diffraction effects and make HCI very challenging \cite{nemati2017effects, pueyo2014high}. In addition to this, thermal distortions due to stellar flux, internal instruments in the observatory, gravitational forces, electrostatic forces, radiation and other observatory dynamics will affect the surface stability of the segmented primary mirror. Each of these factors need to be quantitatively addressed so as to maintain picometer-level wavefront error. 

In this article, we focus on the impact of surface deformations on the segments due to thermal heating of the mounting pads. This thermo-elastic effect associated with the mirror back-plane support structure possesses highest risk to picometer-level wavefront stability and to the performance of the coronagraph. We present a methodology to achieve segment-level thermal stability for the LUVOIR-A architecture. This is required to maximize the yield of discovered and characterized exo-earths. This analysis needs to be included when determining the overall LUVOIR-A wavefront error (WFE) requirements as part of the Ultra-stable Large Telescope Research and Analysis (ULTRA) study\cite{coyle2019large}. We use the finite element models, i.e. the thermal segment models provided by L3Harris Technologies to relate the temperature gradients at the location of the primary mirror to wavefront variations. These models show the localized segment-level surface deformations to 1 mK temperature changes along various directions. In \S  \ref{sec:pastis}, we propagate these custom aberration modes through a diffractive model of the entire LUVOIR-A observatory together with the coronagraphic instrument, and relate the physical quantities (surface deformations and temperature changes) to final the dark-hole contrast. In particular, we use the Pair-based Analytical model for Segmented Telescopes Imaging from Space (PASTIS)\cite{laginja2021analytical,leboulleux2018pair} approach to perform segment-level wavefront error tolerancing in units of physical quantities. The sensitivities obtained from the PASTIS methodology do not include the temporal aspect, and are relevant to the static wavefront control scenario. However, in a real observing scenario environmental conditions drift with time and this induces WFE drifts. The tolerancing statistics (or segment-level sensitivities) required to maintain a desired dark-hole contrast need to be adjusted accordingly. In \S  \ref{sec:temporal}, we translate these sensitivities into the time domain, under the assumption of both open-loop, or ``set and forget", and closed-loop, or continuous wavefront sensing and control (WFS\&C) observing scenarios. We use a model of the LUVOIR-A WFS\&C architecture that measures segment-level errors using a combination of the science camera and a dedicated Zernike out-of-band Sensor. Assuming a perfect controller, we use these sensitivities to relate semi-analytically the open- and closed- loop variance of the segments’ thermo-mechanical modes. We finally tie together the close-loop variance and limiting contrast for each segment-level mode resulting from the finite elements analysis. 

\section{Segment-Level Response to a 1 mK Temperature Change}
\begin{figure}[ht]
\centering
\includegraphics[width=\linewidth]{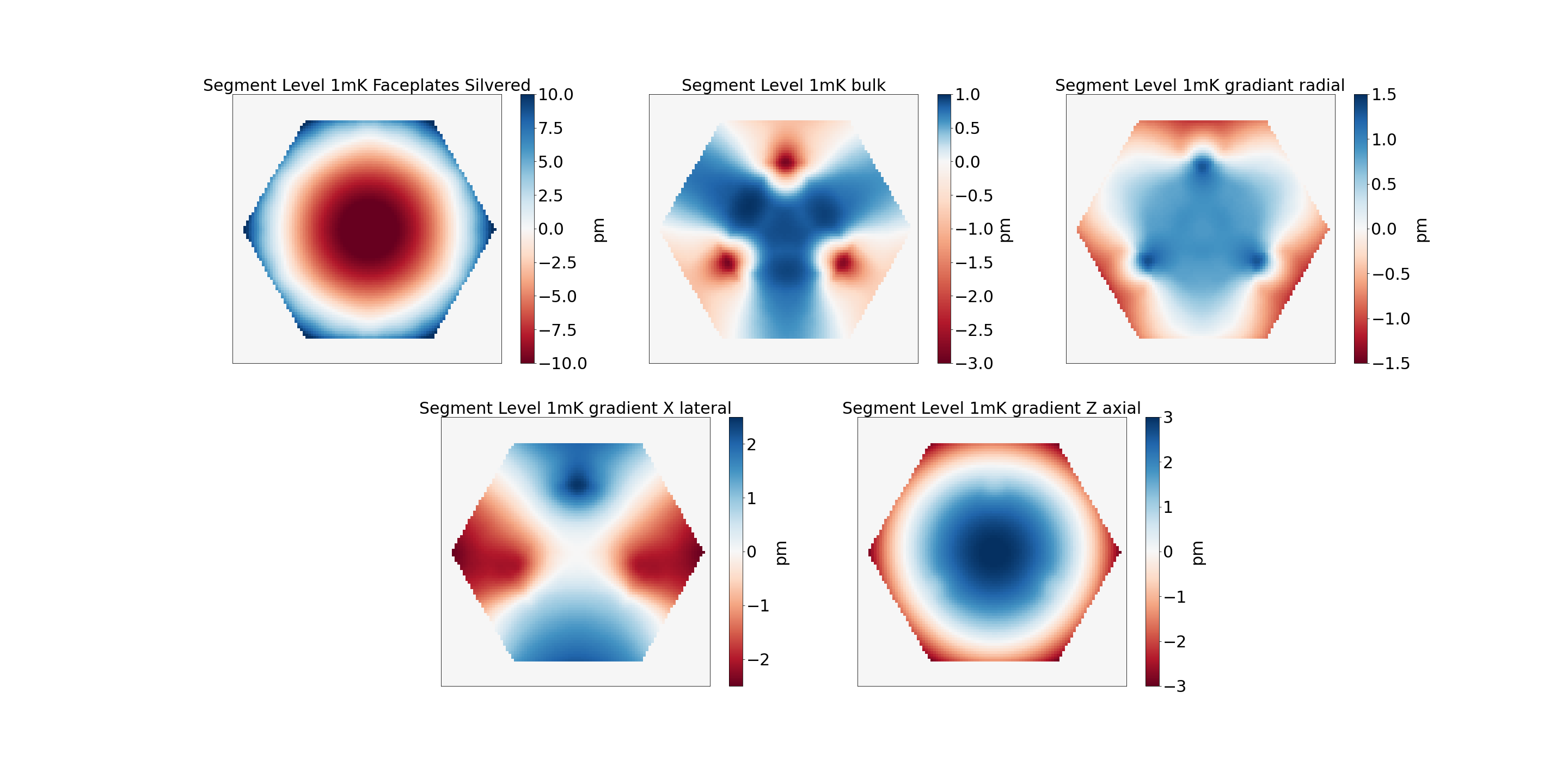}
 \caption{Segment-level surface deformation (or wavefront error maps) due to 1 ~mK temperature change along different directions. \textit{Courtesy: Ball Aerospace and L3Harris Technologies}.} \label{fig:thermal_basis}
\end{figure}

The ULTRA study identifies five kinds of segment-level surface disturbances shown in the Fig.~\ref{fig:thermal_basis} that affect the overall wavefront stability on the picometer level. These disturbances are created when a temperature gradient of 1~mK is applied to a segment along different directions. The mounting pad located underneath the mirror substrate has a high coefficient of thermal expansion and being in direct contact with the mirror induces surface deformation\cite{Coyle2019UltraStableTelescopeResearch} when these gradients are applied. These picometer levels of surface deformations can be measured and actively controlled by the on-board state-of-the-art technologies\cite{saif2017measurement,saif2019picometer}. 
We use these custom segment-level thermal models as our finite elements and propagate them through a diffractive model of the LUVOIR-A architecture. Subsequent sections in this article study the impact of these picometer-level wavefront instabilities on the intensity in the DH of the coronagraphic PSF. In particular, we relate the contrast in the DH to the surface deformation and temperature gradients using the PASTIS sensitivity approach. We inverse this relation to calculate maximum thermal allocation or surface deformation for each segment as required for exo-earth detection.

\section{Using the PASTIS approach for forward propagation} \label{sec:pastis}
Current methods for estimating the DH contrast as a function of the input wavefront aberration involves Monte-Carlo simulations of end-to-end optical propagations with varying numerous factors such as segment phasing errors, surface quality, global vibrations e.t.c.\cite{stahl2013engineering, stahl2015preliminary} Because of the large number of parameters involved, these studies are computationally challenging. Instead, we chose the relatively faster Pair-based Analytical model for Segmented Telescopes Imaging from Space (PASTIS)\cite{laginja2021analytical,leboulleux2018pair} approach to express the focal plane intensity as a quadratic function of the wavefront error in the primary mirror and build a comprehensive error budget in terms of surface deformation and temperature. In the PASTIS framework, the segment-level phase aberration is expressed in terms of Zernike polynomials\cite{lakshminarayanan2011zernike} which are cropped to the support shape of the hexagonal segment. We generalize the PASTIS approach to fit any other basis set of our choice. Instead of regular Zernike polynomials (which are unlikely to occur as localized segment level errors), we use the five segment level thermal maps depicted in Fig.~\ref{fig:thermal_basis} as our basis set. We inverse the PASTIS model to allocate segment-level thermal requirements for a given static DH contrast. 

\begin{figure}[ht]
\centering
\includegraphics[width=0.8\linewidth]{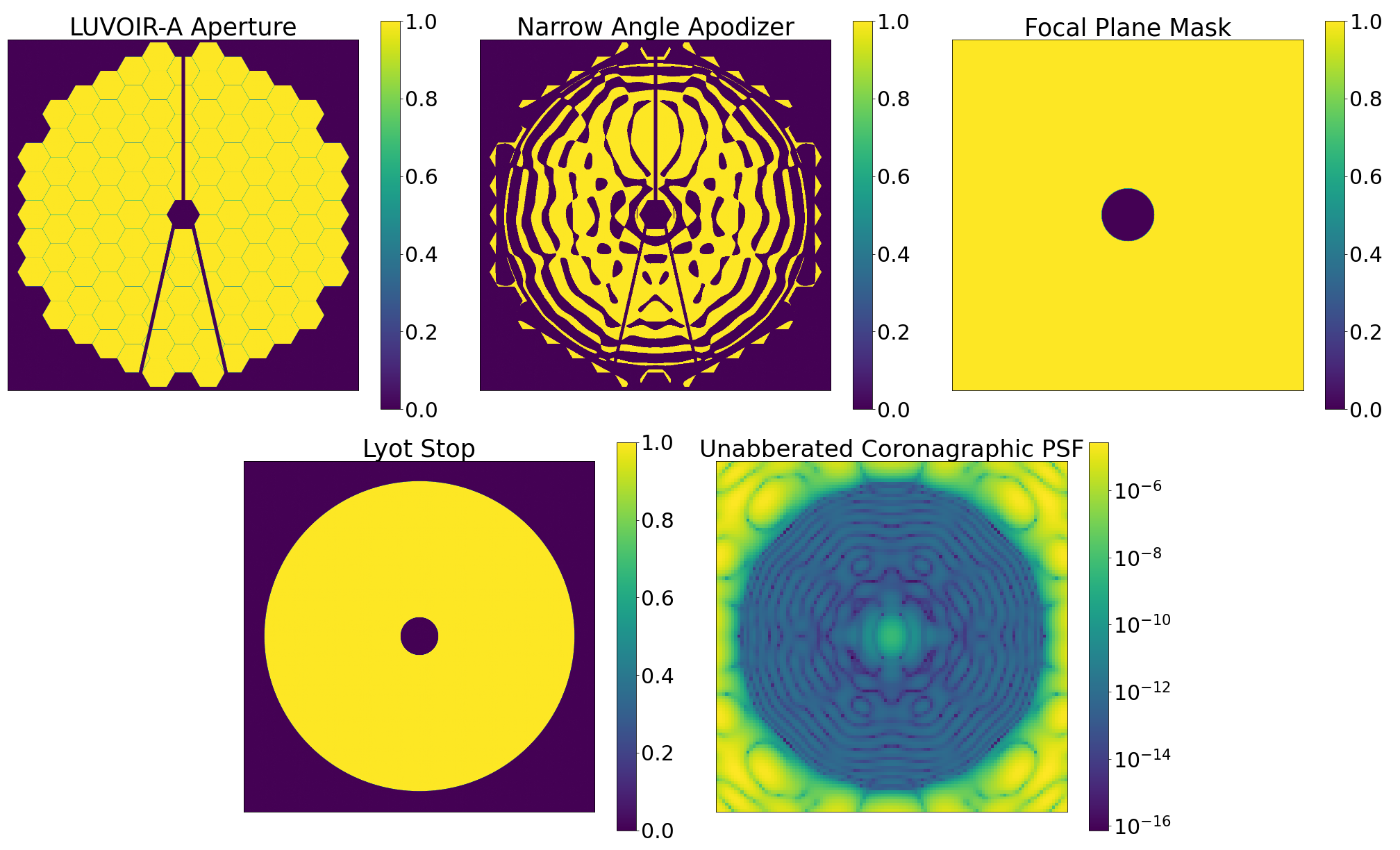}
 \caption{Optical configuration of the LUVOIR-A architecture and the coronagraph instrument. \textit{Top left}: LUVOIR-A primary mirror design with a diameter of 15 m. \textit{Top middle}: Narrow-angle APLC, optimized for exoplanet detection. \textit{Top right}: Focal plane mask of radius $3.5 \lambda/D$. \textit{Bottom}: Lyot stop. \textit{Bottom right}: Simulated focal plane intensity without any wavefront aberration, with a circular DH from $3.4 \lambda/D$ to $12 \lambda/D$. The average contrast $c_{0}$ inside this DH is 4.2$ \times 10^{-11}$.} \label{fig:optical_train}
\end{figure}

The first step to relate the input wavefront aberrations to the DH contrast involves building a model for the forward optical propagation. In this section, we build the semi-analytical PASTIS model using the custom thermal modes. In particular, we build a matrix version for the forward optical propagation, and address it as the ``PASTIS matrix''. Here, we assume a nominal DH has already been created using static coronagraphic mask and dynamic WFS\&C algorithms \cite{pueyo2009optimal, groff2015methods, give2011pair}. Here, we study the influence of a small wavefront perturbation due to thermo-mechanical effects of the primary mirror's back-plane support structure on the spatial contrast in the DH. For a small phase aberration $\phi(\vb{r})$, the electric field $E(\vb{r})$ in the pupil plane can be defined as:
\begin{equation}
\begin{split}
  E(\vb{r}) &= P(\vb{r}) + i\phi(\vb{r}).
\end{split}
\end{equation}
In presence of a coronagraph operator $\mathcal{C}$, the intensity distribution $I_{f}$ in the focal plane can be expressed as:
\begin{equation}
\begin{split}
  {I_{f}}(\vb{u})
  &= |\mathcal{C}\{P(\vb{r})\} + i\mathcal{C}\{\phi(\vb{r})\} |^2  \\
  &= |\mathcal{C}\{P(\vb{r})\}|^2 + 2\mathfrak{Re}({\mathcal{C}\{P(\vb{r})\}\mathcal{C}\{\phi(\vb{r})\}^{*}}) +|\mathcal{C}\{\phi(\vb{r})\}|^2.
\end{split}
\end{equation}
The spatial average of the intensity for a symmetric dark hole is given by:
\begin{equation}
\begin{split}
{\langle I_{f} \rangle}_{DH} &= {\langle{|\mathcal{C}\{P(\vb{r})\}|^2}}\rangle_{DH} + {\langle|i\mathcal{C}\{\phi(\vb{r})\}|^2\rangle}_{DH} \\
 &= c_{0} + {\langle|i\mathcal{C}\{\phi(\vb{r})\}|^2\rangle}_{DH}.
\label{eqn:I-and-c0-averaged}
\end{split}
\end{equation}
Here, $c_{0}$ is the nominal static DH contrast limited by the physical coronagraph and ${\langle|i\mathcal{C}\{\phi(\vb{r})\}|^2\rangle}_{DH}$ is limited by speckles formed due to small wavefront perturbations. The phase $\phi(\vb{r})$ over the entire pupil can be expressed as sum of the segment-level aberrations. Since we are considering only thermal effects, each of the segment-level aberrations can be expressed on a thermal basis \cite{hutterer2018advanced}:
\begin{equation}
    \begin{split}
       \phi(\vb{r}) &= \sum_{(k,l)=(1,1)}^{n_{seg},5} a_{k,l}H_{l}(\vb{r}-\vb{r_{k}}),
       \label{eqn:phi-modes}
    \end{split}
\end{equation}
where $H_{l}(\vb{r}-\vb{r_{k}})$ represents the polynomial form of one of the segment-level surface deformation maps due to 1~mK temperature change, $a_{k,l}$ represents the strength of the aberration in the $k_{th}$ segment for the $H_{l}$ surface deformation map and $r_{k}$ is the coordinate of the center of the $k_{th}$ segment. 
By inserting Eq.~\ref{eqn:phi-modes} into Eq.~\ref{eqn:I-and-c0-averaged}, the average dark-hole contrast $c$ can now be expressed as:
\begin{equation}
    \begin{split}
     c &= c_{0} + {\bigg \langle \bigg| \mathcal{C}\bigg \{ \sum_{(k,l)=(1,1)}^{n_{seg},5} a_{k,l}H_{l}(\vb{r}-\vb{r_{k}})
     \bigg \}\bigg|^2 \bigg \rangle}_{DH} \\
     &= c_{0} + \sum_{(k,l)=(1,1)}^{n_{seg},5}  \sum_{(k',l')=(1,1)}^{n_{seg},5} a_{k,l}a_{k',l'}{\bigg \langle  \mathcal{C}\bigg \{H_{l}(\vb{r}-\vb{r_{k}})
     \bigg \}  \mathcal{C}\bigg \{ H_{l'}(\vb{r}-\vb{r_{k'}})
     \bigg \}^{*} \bigg \rangle}_{DH}.
    \end{split}
\end{equation}
The above equation can be easily written in a matrix form as:
\begin{equation}\label{eqn:simple_matrix}
    \begin{split}
        c = c_{0} + \vb{a}^{T}M\vb{a},
    \end{split}
\end{equation}
where $\vb{a}$ is the pupil-plane aberration represented as a column vector, $M$ is defined as the PASTIS matrix and each of its' element can be defined as follows:
\begin{equation}\label{eqn:matrix_element}
\begin{split}
    m_{k,l,k',l'} &= {\bigg \langle  \mathcal{C}\bigg \{H_{l}(\vb{r}-\vb{r_{k}})
     \bigg \}  \mathcal{C}\bigg \{ H_{l'}(\vb{r}-\vb{r_{k'}})
     \bigg \}^{*} \bigg \rangle}_{DH} \\
     &= {\langle \mathcal{C}\{\vb{e}_{lk} \} \mathcal{C}\{{\vb{e}_{l'k'}} \}\rangle}_{DH},
\end{split}
\end{equation}
where $\vb{e}_{lk}$ is the electric field in the pupil plane associated with the $k^{th}$ segment poke of the primary mirror with the $l^{th}$ thermal mode. Here, the poke amplitude is kept at 1~nm and it can be varied if required. We use an end-to-end diffractive model of the LUVOIR A and a narrow-angle Apodized Pupil Lyot Coronagraph (APLC) to propagate $\vb{e}_{lk}$ and $\vb{e}_{l'k'}$, and compute the complex electric field at the focal plane. A schematic representation of this model is shown in Fig.~\ref{fig:optical_train}. Each element $m_{k,l,k',l'}$ of the matrix $M$ stores the change in the DH contrast resulting due to inferences of two poked electric fields.

\begin{figure}
\centering
\includegraphics[width=0.7\textwidth]{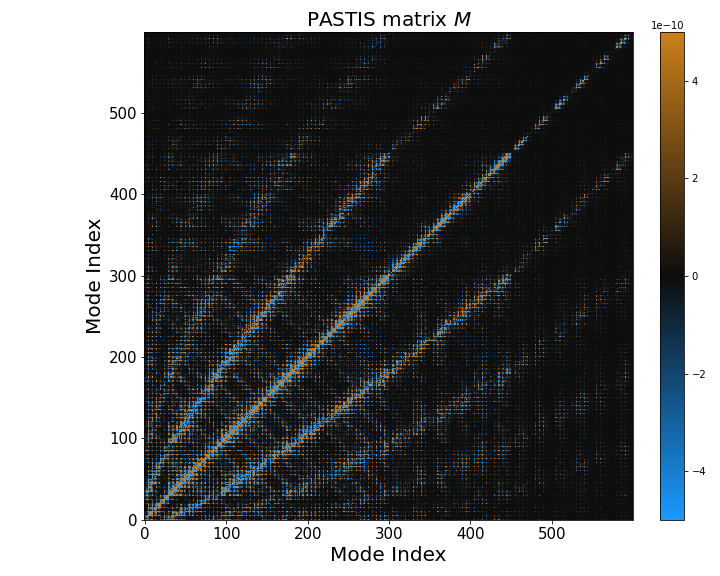}
\caption{PASTIS matrix generated for 600 modes (five distinct basis modes times 120 segments). The diagonal elements have more impact on the contrast than the off-diagonal elements.}
\label{fig:pastisMatrix}
\end{figure}

Figure \ref{fig:pastisMatrix} represents the PASTIS matrix $M$ calculated using Eq.~\ref{eqn:matrix_element} and it relates the final contrast $c$ in the DH to a small-aberration vector $\vb{a}$ using Eq.~\ref{eqn:simple_matrix}. It is a 600$\times$600 symmetric square matrix. The number 600 corresponds to the total number of probed aberration modes: five distinct modes times 120 segments. The nature of the PASTIS matrix is also affected by the design of the coronagraph: the segments that are partially or fully blocked by the APLC have less impact on the final contrast contribution than segments that are more exposed in the pupil. The diagonal elements in this matrix represent the interference terms of two equal electric fields along same direction.They have thus a stronger influence on the contrast compared to the off-diagonal elements.
The statistical mean contrast $\langle c\rangle$ inside the DH for different random aberrations vectors $\vb{a}$ is calculated as follows:
\begin{equation}\label{eqn:covariance}
\begin{split}
    \langle c\rangle &= \langle c_{0} + a^{T}Ma \rangle \\
    & = c_{0} + \text{tr}(M\langle {aa^T} \rangle) \\
    & =c_{0} + \text{tr}(MC_{a}),
\end{split}
\end{equation}
where $C_{a}$ is defined as a  $600 \times 600$ covariance matrix which contains information about the thermo-mechanical correlations between all segments. We note that unlike $C_{a}$, the PASTIS matrix depends on the diffractive model of the observatory and coronagraph. The matrices $M$ and $C_{a}$ together describe the response of the coronagraphic system to thermo-mechanical disturbances.

\begin{figure}[ht]
\centering
\includegraphics[width=\linewidth]{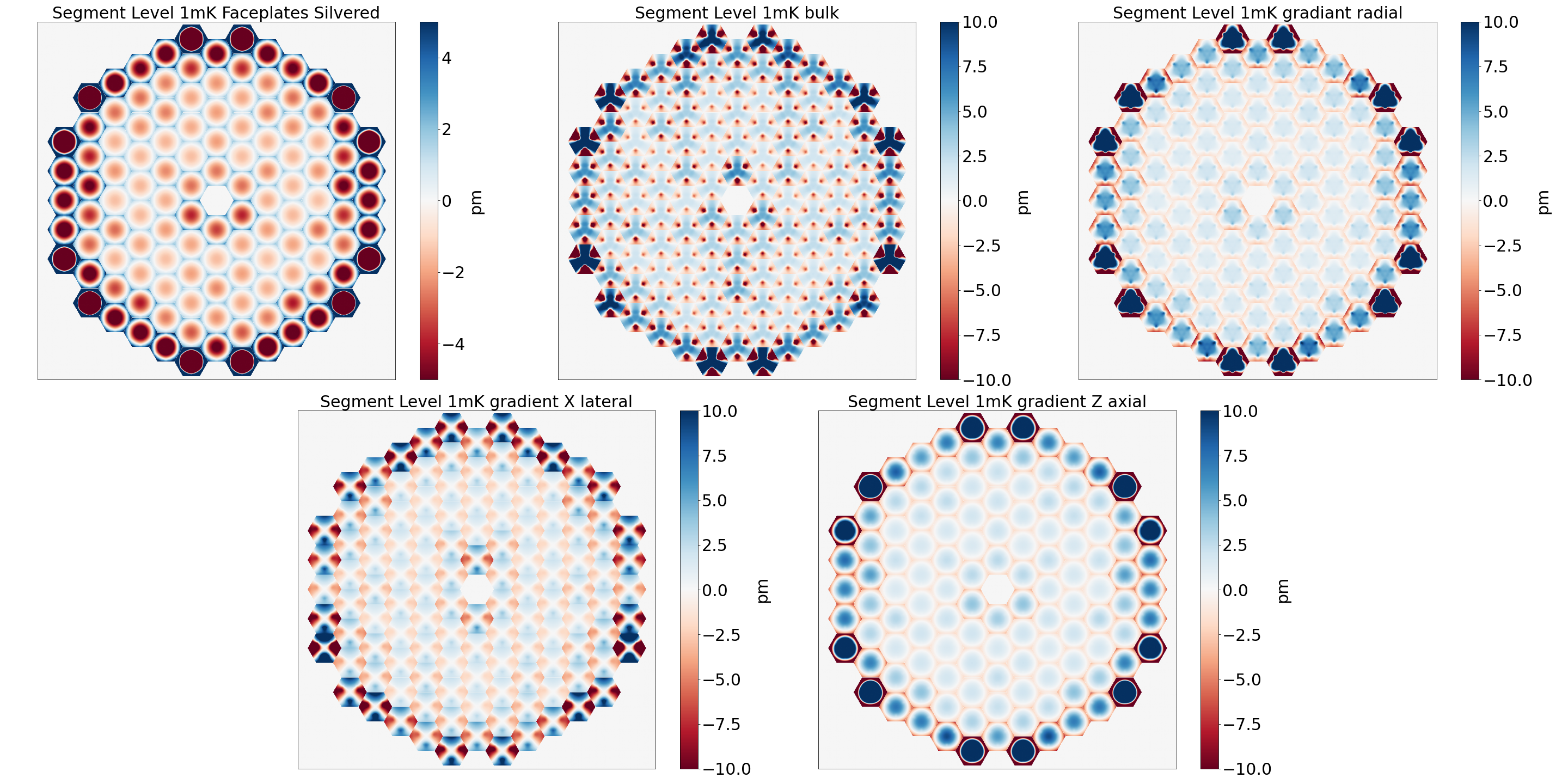}
\caption{Segment-level wavefront tolerance maps required to maintain a mean dark hole contrast of $10^{-11}$ for the narrow-angle APLC at a wavelength of 500~nm. Each segment shows the limit of its surface deformation it can withstand while maintaining the desired contrast. Outer-ring segments are covered more by the APLC apodizer than segments on the inner rings, therefore they have looser tolerance requirements.} \label{fig:mu_map_pm}
\end{figure}

\begin{figure}[ht]
\centering
\includegraphics[width=\linewidth]{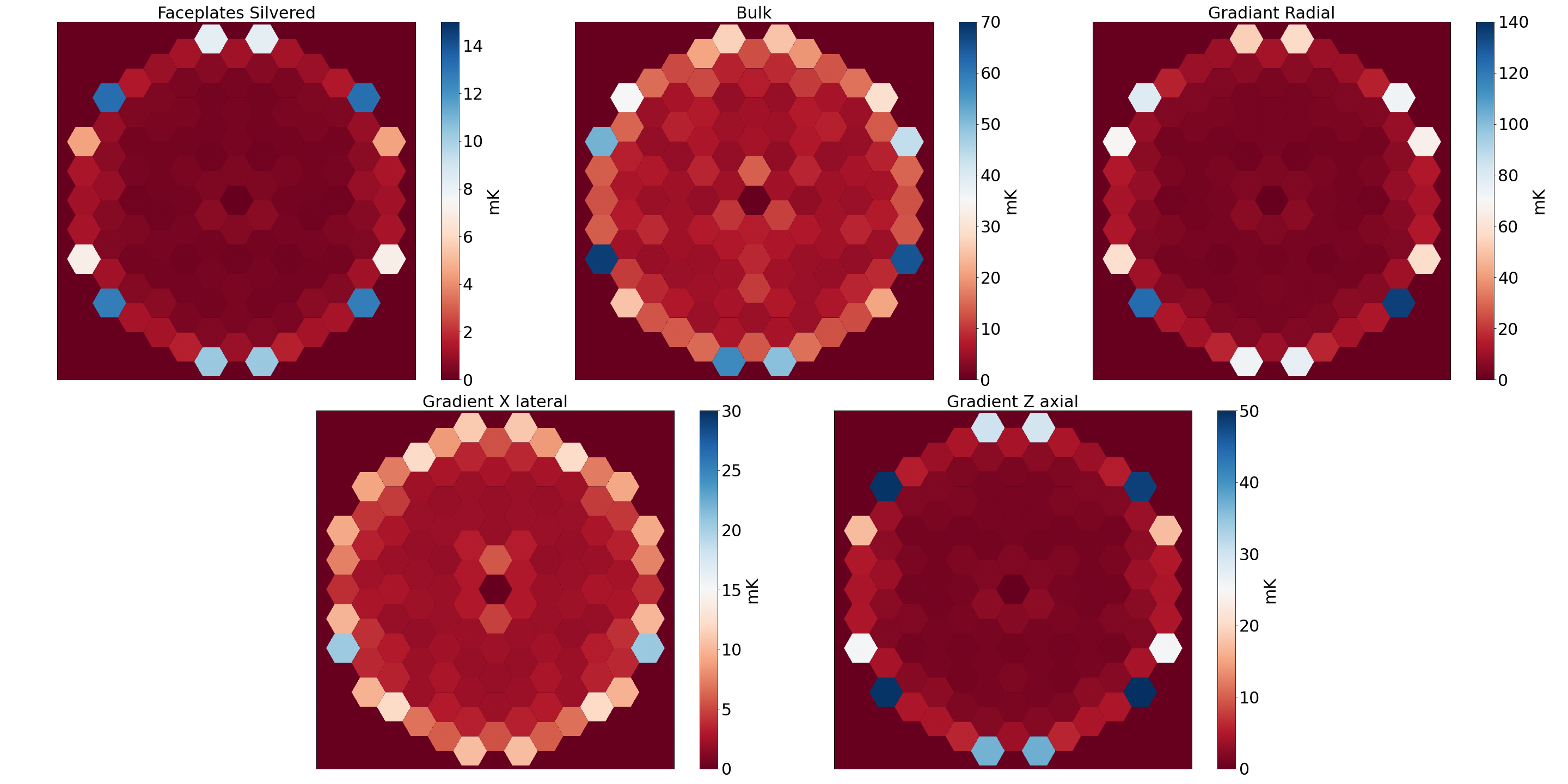}
\caption{Segment Level thermal tolerance maps to maintain a dark hole contrast of $10^{-11}$.The value on each segment represents the the standard deviation for temperature that needs to be applied to each of the segment along a specific direction to satisfy a given DH contrast of $10^{-11}$, for a narrow angle APLC at a wavelength of 500 nm.} \label{fig:mu_map_mK}
\end{figure}

For simplicity, we assume all the segments are independent of each other and we have access to the statistical mean of the spatial average contrast in the DH. In this scenario, $C_{a}$ will be a diagonal matrix and Eq.~\ref{eqn:covariance} is simplified:
\begin{equation}\label{eqn:mu_k}
\begin{split}
    \langle c\rangle &= c_{0} + \sum_{k = 1}^{600} m_{kk}\langle{a^2_{k}}\rangle \\
    &= c_{0} + \sum_{k = 1}^{600} m_{kk}\mu^2_{k},
\end{split}
\end{equation}
where $m_{kk}$ is the diagonal element of the PASTIS matrix, $a^2_{k}$ is the aberration amplitude and $\mu^2_{k}$ is the standard deviation of the wavefront aberration on $k_{th}$ segment. In the case where each mode contributes uniformly to the DH contrast, Eq.~\ref{eqn:mu_k} is reduced to:
\begin{equation}
    \langle c\rangle = c_{0} + 600\times m_{kk}\mu^2_{k}
\end{equation}
and 
\begin{equation}\label{eqn:mu_k2}
    \mu_{k} = \sqrt{\frac{\langle c\rangle - c_{0}}{600\times m_{kk}}}.
\end{equation}
Eq.~\ref{eqn:mu_k2} allows us to calculate the required variances per thermal mode, per segment, for all individual segments of the primary mirror in order to maintain a given statistical mean contrast inside the DH over many realizations of wavefront error maps built by the thermal basis set. Fig.~\ref{fig:mu_map_pm} shows the segment level tolerance requirements to main a DH contrast of $10^{-11}$. It contains both the information about the thermal mode number ($l$ index) and the segment number ($k$). The depth of color on each segment represents the strength of the permissible wavefront error as a standard deviation. 
We recall that the five thermal basis maps were generated by applying 1~mK temperature change to a segment along various directions. Each point on the five thermal maps shown in Fig.~\ref{fig:thermal_basis} represents the surface deformation (in pm) per 1~mK temperature change. Therefore, we divide each of these segment-level surface requirement maps in Fig.~\ref{fig:mu_map_pm} with their corresponding thermal modes to obtain the required heat limit per each segment. The segment-level thermal tolerance maps shown in Fig.~\ref{fig:mu_map_mK} depict the maximum standard deviation temperature that needs to be applied to each of the segments, along a specific direction, to satisfy a mean DH contrast of $10^{-11}$. Table \ref{table:1} summarises the maximum, average and minimum temperature required across all segments to maintain a DH contrast of $10^{-11}$.
\begin{table}
\centering
\caption{Segment-level temperature requirements to maintain a DH contrast of $10^{-11}$ with a narrow-angle APLC at a wavelength of 500~nm.\\}
\label{table:1}
\begin{tabular}{|p{4cm}|p{3cm}|p{3cm}|p{3cm}|}
 \hline
 Mode Index & Min (in mK) & Max (in mK)& Average (in mK)\\
 \hline
 Faceplate Silvered & 0.27 &13.18& 1.50\\
 Bulk & 4.07 & 67.86 & 11.44\\
 Gradient Radial &2.62 & 135.25&13.16\\
 Gradient X Lateral &1.84& 20.51&4.45\\
 Gradient Z Lateral &0.98& 50.65&5.52\\
 \hline
\end{tabular}
\end{table}

\section{Temporal analysis to limit and correct thermal drifts}\label{sec:temporal}
Small thermal and mechanical disturbances in the observatory conditions can significantly affect the incoming stellar wavefront and the performance of the coronagraph used in the system. These disturbances lead to dynamic speckles in the image plane and degrade the raw contrast inside the dark hole. Subsequently, new tolerance maps are required to compensate the drift and maintain a desired stable contrast inside the dark hole. In \S~\ref{sec:pastis}, the PASTIS approach is limited to defining only the static tolerances for the segments and it does not specify in what way these tolerances are to be maintained during a long science observing scenario. In this section, we discuss drifts associated with the localized spatial wavefront error on all segments, and apply the algorithm established in Pogorelyuk et al. 2021\cite{pogorelyuk2021information} to maintain these dynamic drifts through a continuous wavefront sensing and control strategy. This algorithm applies to a scenario where a nominal dark hole has already been created using static coronagraphic masks and dynamic WFS\&C algorithms such as the pair-wise estimation and the electric field conjugation (EFC) methods.\cite{give2011pair, thomas2010laboratory}

\begin{figure}[ht]
\centering
\includegraphics[width=0.8\linewidth]{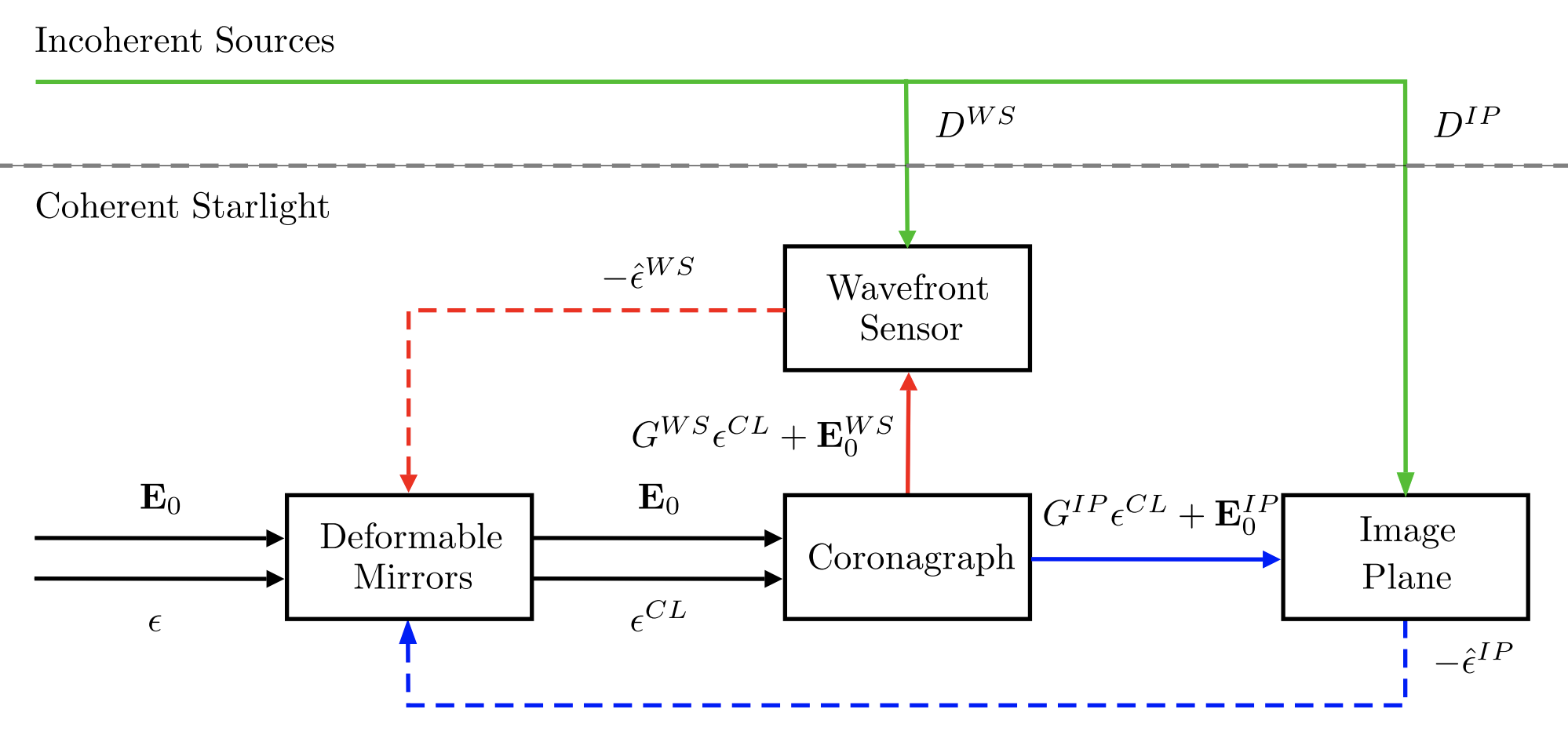}
\caption{A schematic representation of the AO system used to correct the WFE $\epsilon$ arising due to thermal drifts. The tolerance maps generated in the previous section are treated here as open-loop coefficients of the WFE modes $\epsilon$. These coefficients are partially corrected by the DM based on the estimated WFE (i.e., $\epsilon^{WS}$) by the wavefront sensor and the WFE estimated at the image plane (i.e., $\epsilon^{IP}$). \textit{Courtesy: Pogorelyuk et al. 2021\cite{pogorelyuk2021information}}} \label{fig:ao_sys}
\end{figure}

\begin{figure}[ht]
\centering
\includegraphics[width=0.7\linewidth]{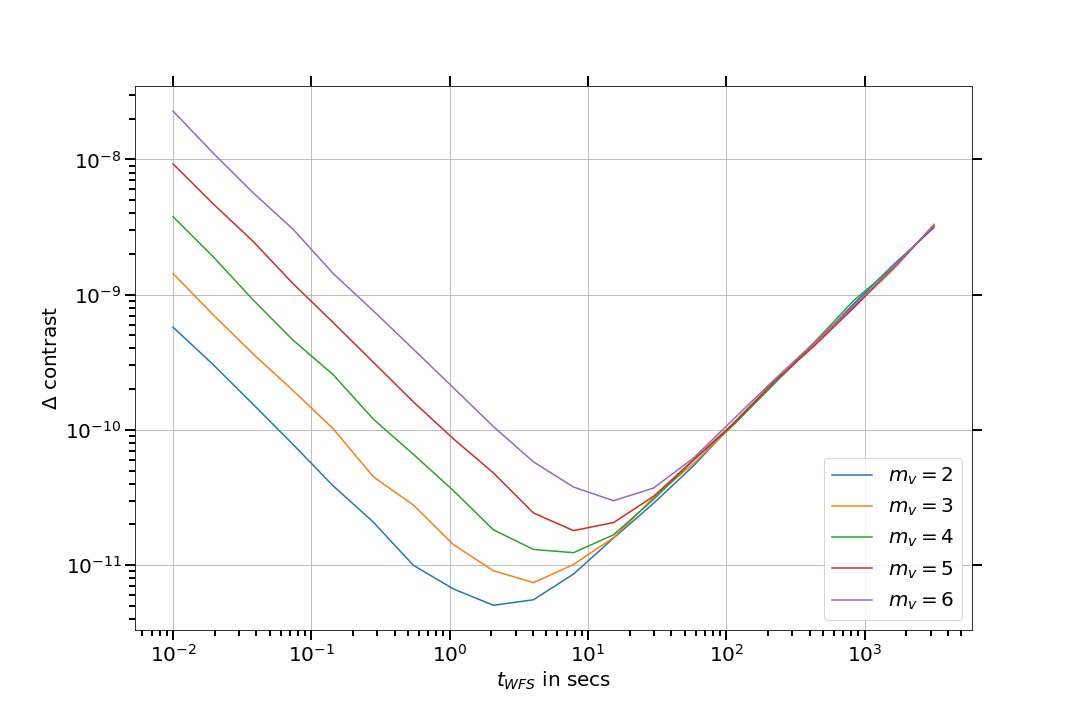}
\caption{Estimated change in image plane contrast as a function of wavefront sensor exposure time for different stellar magnitudes. We have assumed no detector noise for this case.} \label{fig:cont_mv}
\end{figure}

\begin{figure}[ht]
\centering
\includegraphics[width=0.7\linewidth]{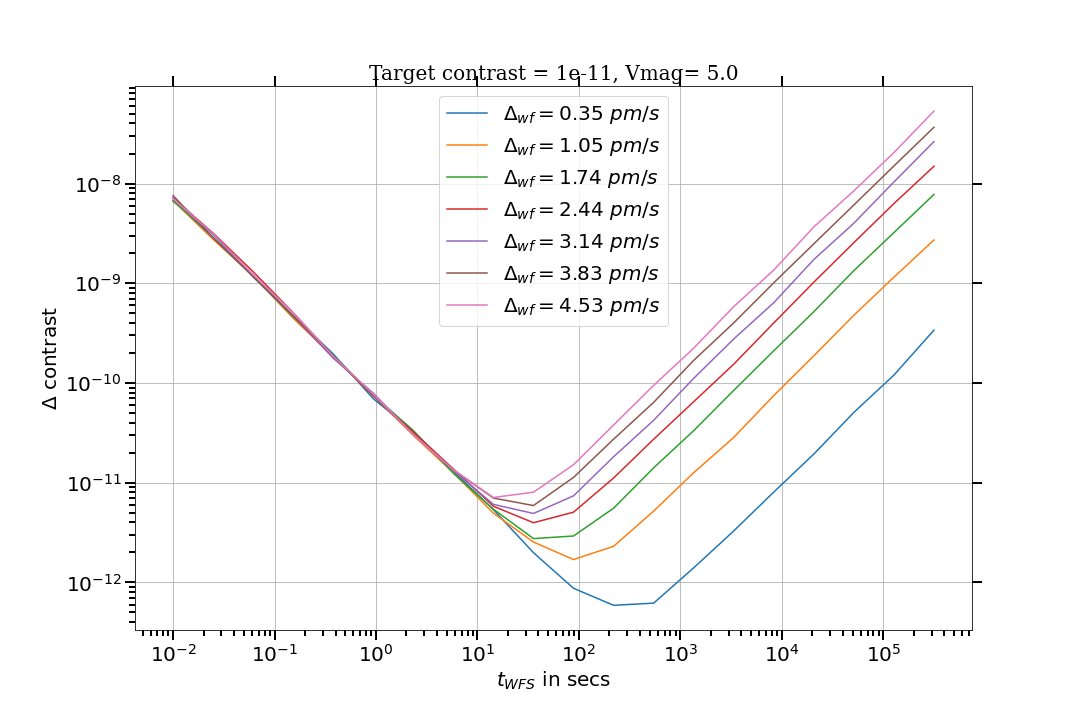}
\caption{Estimated change in image plane contrast as a function of wavefront sensor exposure time for different amplitudes of wavefront aberration. This plot is for a 5th magnitude star in the visible band. We have assumed no detector noise for this case.} \label{fig:cont_wf}
\end{figure}

\begin{figure}[ht]
\centering
\includegraphics[width=\linewidth]{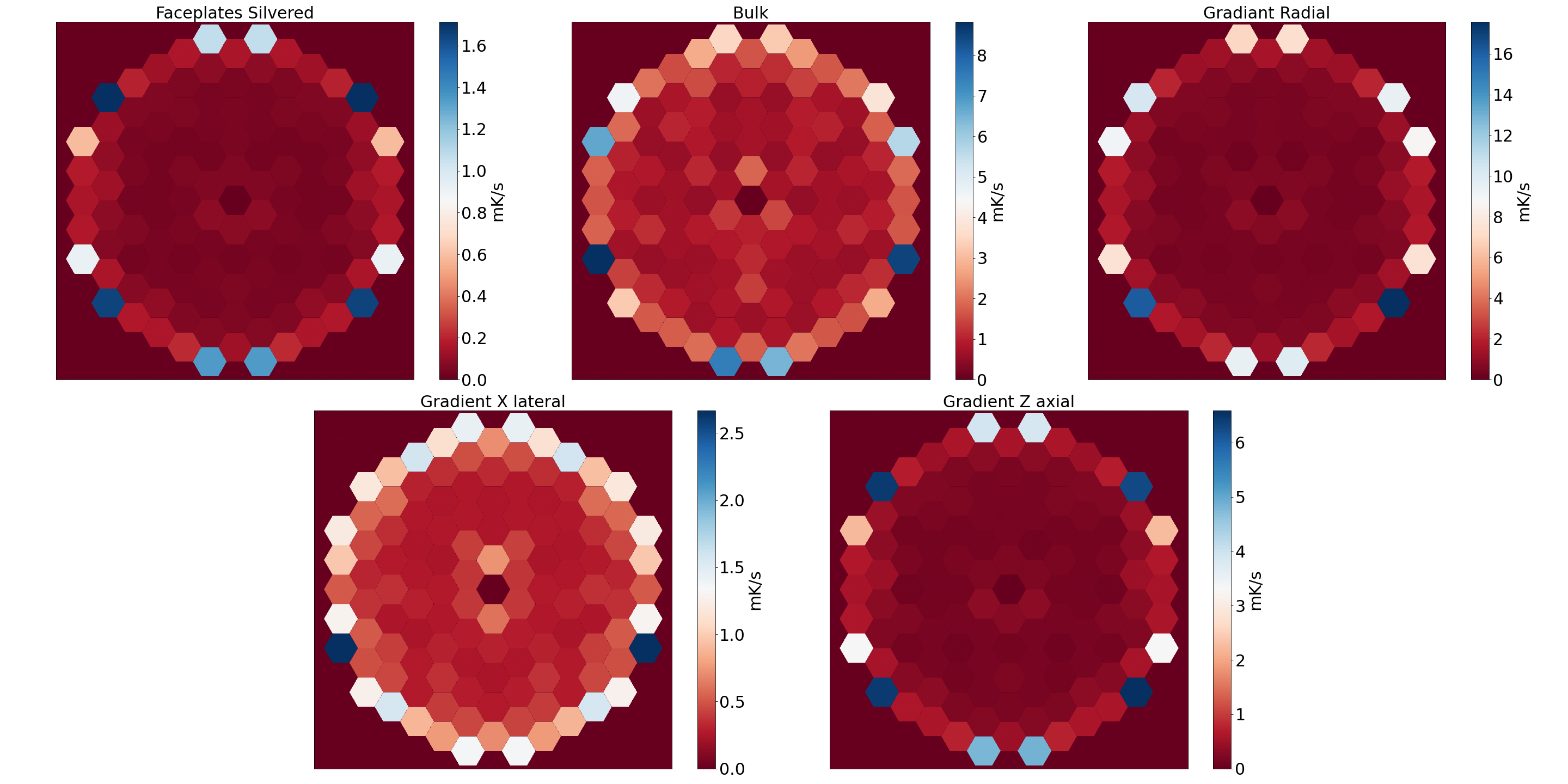}
\caption{Closed-loop thermal sensitivity maps to maintain a DH contrast of $10^{-11}$ for a 5th magnitude star in visible at a wavelength of 500 ~nm. These plots are generated by obtaining the amplitude aberration from Fig.~\ref{fig:cont_wf}.} \label{fig:mu_map_mKs}
\end{figure}

We estimate a lower bound on the wavefront variances based on the Cramer-Rao inequality\cite{cramer1946contribution, rao1992information}, and then relate these variances to the residual starlight intensity in the dark hole region of coronagraphic PSF. The algorithm is used to maintain a small and slow drift of wavefront aberrations during a scientific observation. Let $\epsilon$ be the coefficients for wavefront errors at $k_{th}$ exposure, and let increments of these coefficients be normally distributed with:
\begin{equation}
    \epsilon_{k+1} - \epsilon_{k} \sim \mathcal{N}(\textbf{0}, Q), \; Q>0,
\end{equation}
where $Q$ is the drift covariance. Let $\hat{\epsilon}$ be the \textit{unbiased} or the perfect estimate of the coefficient of the wavefront error mode, and its error is also normally distributed with a covariance $P_{k}$, i.e.:
\begin{equation}
    \hat{\epsilon_{k}} - \epsilon_{k} \sim \mathcal{N}(\textbf{0},P_{k}),       \; P_{k} > 0.
\end{equation}
The closed-loop estimate of the coefficients of the WFE modes $\epsilon^{CL}$ is expressed as:
\begin{equation}
\begin{split}
\epsilon^{CL}_{k+1} = \epsilon_{k+1}-\hat{\epsilon_{k}}.
\end{split}
\end{equation}
In the case of an imperfect controller or estimator, $\epsilon^{CL}$ is a non-zero quantity, and is also normally distributed:
\begin{equation}
    \begin{split}
        \epsilon^{CL}_{k+1} \sim \mathcal{N}(\textbf{0}, P_{k} + Q).
    \end{split}
\end{equation}
We also assume that the electric field $E^{IP}$ at the image plane is a linear function of the closed-loop drift $\epsilon^{CL}$ and is expressed as:
\begin{equation}
    E^{IP} = G^{IP}\epsilon^{CL} + E^{IP}_{0},
\end{equation}
where $G^{IP}$ is a sensitivity matrix at the image plane and $E^{IP}_{0}$ is a static, uncontrolled reference electric field without any aberrations. Here, $G^{IP}$ holds information about the electric fields to the known wavefront aberrations. The intensity at the image plane including photon flux $\dot{N_{S}}$ from the source can further be written as:
\begin{equation}
I = \dot{N_{S}}||G^{IP}\epsilon^{CL} + E^{IP}_{0}||^2.
\end{equation}
From $I$, we compute the probability distribution for the measured number of photons $y$ as:
\begin{equation}
    \text{pmf}(y) = \frac{1}{y!}((I+D^{ext}+D^{int})t_{s})^{y}e^{-(I+D^{ext}+D^{int})}t_{s},
\end{equation}
where $D^{ext}$ and $D^{int}$ refer to photons from external sources (such as zodiacal dust) and internal sources (clock-induced charge, dark current), respectively. Unlike $\text{pmf}(y)$ which can be directly estimated, we use The Fisher information approach to estimate indirectly the closed-loop WFE modes $\epsilon^{CL}$. The Fisher information $\mathcal{I}$ relate $\epsilon^{CL}$ to the intensity, either at the image plane or at the wavefront sensing plane through the Cramer-Rao inequality and is expressed as:
\begin{equation}\label{eqn:fisher}
\begin{split}
    \mathcal{I} &= {\Bigg \langle{ \left( \frac{\partial \log \text{pmf}(y)}{\partial \epsilon^{CL}} \right)\left(\frac{\partial \log \text{pmf}(y)}{\partial \epsilon^{CL}}\right)^{T}} \Bigg\rangle}_{y}  \\
     & = \frac{4\dot{N_{S}}t_{s}}{||G^{IP}\epsilon^{CL} + E^{IP}_{0}||^2+N^{-1}_{S}(D^{ext}+D^{int})}G^{T}(G^{IP}\epsilon^{CL} + E^{IP}_{0})(G^{IP}\epsilon^{CL} + E^{IP}_{0})^{T}G.
\end{split}
\end{equation}
Here, we are interested in estimating a lower bound on $P$, (the closed-loop variance of the wavefront error mode coefficients). It is also assumed that $P$ does not change much for the nearest exposure, meaning that ($P_{k+1}\approx P_{k}\approx P$). In this scenario, the Fisher information can be approximated as:
\begin{equation}
    \mathcal{I}_{k+1} \approx \mathcal{I} \approx 
    \langle\mathcal{I}\rangle_{\epsilon^{CL}}, \forall \epsilon^{CL} \in \mathcal{N}(\textbf{0}, P+Q) .
\end{equation}
We use the following iterative batch estimation algorithm to compute a steady-state WFE covariance estimate $P$: 
\begin{itemize}
  \item We initialize $P_{k}=0$
  \item We draw a random sample of $\epsilon_{k+1}^{CL}$ from the a normalized distribution having a covariance of  $P_{k}+Q$, and calculate $\mathcal{I}$ based on the intensity in the image plane and wavefront sensor plane.
  \item After each iteration, we update $P_{k+1} = {I_{k+1}}^{-1}$
  \item The above two steps are repeated until $P$ reaches an arbitrarily small value.
\end{itemize}
We apply the algorithm introduced to get an estimate on the steady state closed-loop wavefront error. A schematic architecture of the used AO system is given in Fig.~\ref{fig:ao_sys}. The tolerance maps (see Fig.~\ref{fig:mu_map_pm}) that were generated in \S .~\ref{sec:pastis} are treated here as the covariance $Q$ of the open-loop WFE modes $\epsilon$. We propagate $Q$ through the AO system. These coefficients are partially corrected by the deformable mirror (DM) using the estimated WFE (i.e., $\epsilon^{WS}$) by the wavefront sensor, and the WFE estimated at the image plane (i.e., $\epsilon^{IP}$). In the wavefront sensing plane, we use an out-of-band Zernike wavefront sensor to estimate $\epsilon^{CL}$. We assume that a nominal dark hole has already been created by the primary extreme AO loop and we are looking at the impact of small perturbations $\epsilon$ on the contrast. In this regime, we assume that the image-plane electric field is a linear function of the sensitivity matrix $G^{IP}$ and the incoming electric field $E_{0}$. We calculate the Fisher Information based on the expression in Eq.~\ref{eqn:fisher} and update $P$ until goes to an arbitrarily small values. Figure \ref{fig:cont_mv} shows the change in closed-loop contrast (i.e., dark-hole contrast of the abberated PSF - dark-hole contrast of the reference PSF) at different wavefront sensing time scales. A fainter target takes a longer time to reach the minimum dark-hole contrast. On the left-hand side of the minima, the contrast is limited by photon noise, and on the right-hand side it is dominated by speckle noise so the contrast remains the same irrespective of the stellar magnitude. We have assumed no detector noise for this case.
Figure \ref{fig:cont_wf} represents the change in closed-loop contrast as a function of wavefront sensor exposure time for different amplitudes of open-loop WFE drifts $\epsilon$. From this plot, we obtain an optimal WFS exposure time of $\sim~25$~s to reach the dark-hole contrast of $10^{-11}$ for an open-loop WFE drift of 4.53~pm/s across all five thermal modes. These parameters are valid for a 5th magnitude star in the visible, and might change for other magnitudes. Figure \ref{fig:mu_map_mKs} shows how much thermal drift we need to correct each second such as to maintain a DH contrast of $10^{-11}$ for a 5th magnitude star. 

\section{Summary}
We studied the impact of thermo-elastic effects of the mirror mounting pads of the primary mirror on the final dark-hole contrast in the image plane. We used the PASTIS forward propagation  to relate surface deformations of the segments to the dark-hole contrast. We set tolerances both in terms of surface deformation and temperature gradient for each of the 120 segments of the LUVOIR-A architecture with a narrow-angle APLC. Assuming a uniform contrast allocation (i.e., each segment independently contributes equally to the total dark-hole contrast), we find that the inner segments are to be constrained more tightly than the outer ones. During a long science observation, the environment condition of the observatory is expected to drift. We implemented a closed-loop batch estimation algorithm to measure and maintain the wavefront drifts in a real science observation scenario. For a fifth magnitude bright star in the visible, with zero detector noise, we demonstrated that we can correct temperature drifts to at least of an order of 1~mK per sec. Similar to the static tolerance analysis, we see that the outer segments are temporally loosely constrained compared to the inner ones. Future studies will include extending this analysis to other segment-level mechanical aberrations, orientations of the mounting pads and tolerances at different segment numbers, temporal scales and higher order spatial wavefront aberrations. 
\acknowledgments
This work was co-authored by employees of BALL AEROSPACE as part of the the Ultra-Stable Telescope Research and Analysis (ULTRA) Program under Contract No. 80MSFC20C0018 with the National Aeronautics and Space Administration (PI: L. Coyle), and by STScI employees under corresponding subcontracts No.18KMB00077 and No. 19KMB00102 with Ball Aerospace (PI: R. Soummer, Sci-PI: L. Pueyo). This work was supported in part by the NASA Grant \#80NSSC19K0120 issued through the Strategic Astrophysics Technology/Technology Demonstration for Exoplanet Missions Program (SAT-TDEM; PI: R. Soummer) and in part funded by the STScI Director’s Discretionary Research Fund. The authors wish to acknowledge the critical importance of current and recent research staff, scientist, astronomers, computer support, office employees at the Space Telescope Science Institute, Baltimore, MD  for their expertise, ingenuity and dedication to the development of future space missions. IL acknowledges the support by a postdoctoral grant issued by the Centre National d'Études Spatiales (CNES) in France. 
  
\bibliography{report} 
\bibliographystyle{spiebib} 
\end{document}